# Batch versus microfluidic emulsification processes to produce whey protein microgel beads from thermal or acidic gelation

Alban LACROIX[a], Murielle HAYERT[a], Veronique BOSC[a], Paul MENUT[a]

[a]Université Paris-Saclay, INRAE, AgroParisTech, UMR SayFood, 91077, Massy, France
Corresponding author: Paul MENUT, email: paul.menut@agroparistech.fr

**Highlights:**
- Food-grade spherical microgels can be formed from either microfluidic or batch emulsification.
- Polydispersity was ten times lower for particles produced with microfluidics.
- Acid gelation gave homogeneous spherical microgels.
- Thermal gelation resulted in aggregated particles or in particles with an irregular surface.

**Abstract:**
*Producing food-grade soft particles with controlled structure is of interest to elucidate the structure-properties relationship in soft-particles suspensions. The aim of this work is to evaluate the ability of two elaboration processes to produce homogenous and spherical whey protein microgels with adjustable diameters in the range 40-100 µm. Microgels are formed in two steps: (1) emulsification of a whey protein aqueous solution in oil and (2) gelation of the protein solution droplets. We compare a continuous emulsification in a home-made microfluidic device, designed on purpose, with a more simple emulsification by mixing. In addition, two gelation processes are studied: a thermal gelation at 80°C and an acid gelation. Results s̶show that emulsification controls the size polydispersity (pdI<0.1 for microfluidics) while gelation controls the microgels structure and assembly. Acid gelation in the microfluidic device results in spherical, homogeneous microgels which properties are controlled by the process parameters.*

# 1 Introduction

Soft particles such as gelatinised starch granules, plant cells or dairy microgels play an important role in the textural and sensorial properties of a large variety of food suspensions such as thickened sauces, purees or stirred yoghurt (Espinosa-Muñoz et al., 2012; Szczesniak, 2002; Wilkinson et al., 2001). Rheological properties depend on both the suspension characteristics, such as the particles volume fraction and continuous phase viscosity, and on the structural characteristics of the particles themselves, such as their size, shape and rigidity (Faroughi and Huber, 2017; Menut et al., 2012). In food, such particles are built from natural, food-grade biopolymers, and usually show a responsive behaviour under changes in pH, ionic strength, osmotic pressure or temperature. Production of suspensions of particles with specific characteristics (e.g. size or softness) would help to determine the precise effect of this single particle characteristic on the whole textural properties. Two approaches are possible to produce such suspensions: a top-down and a bottom-up approach. In the top-down approach, an already existing suspension is processed to modify the particles properties. For example reducing the average diameter in a plant cell based suspension can be done by grinding and/or sieving (Leverrier et al., 2017). However, this transformation usually affects simultaneously different characteristics, for example, reducing the particle size often comes with changes in polydispersity (Andoyo et al., 2016; Leverrier et al., 2016). By contrast, in the bottom-up approach, particles are formed from individuals building blocks (such as proteins or carbohydrates) to obtain the desired, predefined properties. This can theoretically give a better control over particles and suspension properties.

Microgels are microscopic-size hydrogels, which intrinsic properties are similar to their macroscopic counterpart. They are formed using either a physico-chemical or a physical method (Farjami and Madadlou, 2017; Fernandez-Nieves et al., 2011; Shewan and Stokes, 2013). Physico-chemical methods imply the fine tuning of physico-chemical interactions in solution to assemble building blocks into soft structures, while physical methods combine physical and mechanical stresses to tailor microgel preparation.

Physico-chemical methods include coacervation or phase separation. In defined conditions, electrostatic interactions, steric exclusion or depletion interactions lead to the spontaneous formation of liquid droplets of a concentrated phase, in equilibrium with a diluted phase in which it is dispersed (Boire et al., 2019) that will later on gel (Chen et al., 2020). This process depends on physico-chemical parameters such as pH, salt, temperature, concentration or anion-cation interactions, that control the biopolymers interactions (Perfetti et al., 2018, 2020). Recently, it was also shown that the dry heating directly applied on a protein powder could be used to produce protein microparticles that can later on swell in water to give microgel particles (Famelart et al., 2021, 2018).

Physical methods can consist in the simple shearing of a macroscopic gel, or in droplet based techniques. In the first option, the shear stress can be applied either during or after gelation. Beside relatively simple, this method usually gives microgels with irregular shapes and size (characterized by a large polydispersity), which depend on the applied shear stress (Jones and McClements, 2010; Leon et al., 2016; Norton et al., 1999; Young et al., 2021) and gel mechanical properties (Saavedra Isusi et al., 2019). In the droplets based technics, droplets including the gel precursors are initially formed before gelation. Owing to the spherical nature of droplets, this allows the production of spherical microgels after droplets gelation, but also the fine control of the microgel size, which depends on the droplets diameter. Droplets are therefore a perfect template for spherical microgel preparation. They can be formed directly by the fractionation of the precursor solution in the air, for which different technics are available, such as spinning disk atomisation (Hilborn, 1995; Senuma et al., 1999), extrusion (Burey et al., 2008; Prüße et al., 2000; Torres et al., 2017) or spray drying (Burey et al., 2008; Perrechil et al., 2011). These technics allow large scale microgel production (Shewan and Stokes, 2013). The resulting microgels are usually monodispersed but less spherical than emulsion based technics (Kuhn et al., 2019; Marra et al., 2017; Perrechil et al., 2011).

By contrast, emulsification methods are more easy to use at a laboratory or pilot scale. Different methods are listed in the literature:
- Batch emulsification : the aqueous phase is poured into an oil phase while stirring the solution (Andoyo et al., 2016; Sağlam et al., 2014);
- Microfluidics : the aqueous phase is pumped into a laminar flow of oil (Heida et al., 2017; Marengo et al., 2019; Santos et al., 2020);
- Membrane emulsification : the aqueous phase is pushed through a membrane into an oil phase (Maleki et al., 2021; Tran et al., 2011).

Batch emulsification allows the formation of droplets with sizes ranging from a few microns to hundreds of microns, by varying the stirring speed during the emulsification process (Andoyo et al., 2016; Dingenouts et al., 1998; Sağlam et al., 2011). Microfluidic is a more recent technic, which offers the possibility of a fine control of droplets generation (Zhang et al., 2020), the most versatile geometry being the flow-focusing one (Costa et al., 2017; Marengo et al., 2019). Monodisperse droplets with sizes ranging from tens to hundreds of microns can be produced, their size depending on the size of the microfluidic channels and on the flow ratio between the dispersed and continuous phase (Shimanovich et al., 2014).

Food-grade microgels can be prepared from a water-soluble food biopolymer such as gelatine (Burey et al., 2008), alginate (Damiati, 2020), carrageenan (Burey et al., 2008; Marengo et al., 2019) Rodriguez et al., 2021), whey proteins (Andoyo et al., 2016) or pea proteins (Ben-Harb et al., 2018). Whey Protein Isolates (WPI) are interesting precursors for gelation because of their high solubility in water and the variety of gelation methods that can be used (Andoyo et al., 2016; Foegeding et al., 2002; Ikeda, 2003; Z. Y. Ju and Kilara, 1998a; Z.Y. Ju and Kilara, 1998; Langton and Hermansson, 1992; Mcclements and Keogh, 1995). For a high enough concentration, WPI can form a gel through two different processes: thermal gelation or addition of a geling agent (usually called cold-set gelation as it does not request heating). In the case of thermal gelation (Horne et al., 2001; Z. Y. Ju and Kilara, 1998a; Morr and Foegeding, 1990), the solution is heated above 80 °C to ensure thermal denaturation: proteins unfold and new inter-proteins bonds are created, that eventually result in the formation of a percolating network. The addition of a gelling agent (a salt, an acidifier or an enzyme) induces gelation of proteins that were usually previously denatured. The addition of salt reduces electrostatic repulsion so that proteins aggregate. Large protein aggregates then quickly form a network, later join by smaller aggregates (Z.Y. Ju and Kilara, 1998; Marangoni et al., 2000; Mcclements and Keogh, 1995). Controlled acid gelation can be obtained by Glucono-delta-Lactone (GDL) addition (Z. Y. Ju and Kilara, 1998b; Kharlamova et al., 2018). GDL hydrolyses and forms gluconic acid. As the pH decreases and approaches the isoelectric point (pHi), electrostatic repulsions are reduced and denatured proteins form a network. The kinetic of gelation is here controlled by the GDL to protein ratio (Cavallieri and da Cunha, 2008). Enzymatic gelation is obtained with the addition of an enzyme, for example the *Bacillus Licheniformis Protease* (Ju et al., 1997; Spotti et al., 2017) which hydrolyses proteins and allows the formation of hydrogen bonds, or *Transglutaminase* (Liang et al., 2020), an amino-transferase which is responsible of the formation of covalent bonds between proteins through lysine glutamine residues. These different processes give a diversity of WPI gel thus a diversity of gel mechanical properties (Foegeding et al., 2002; Z. Y. Ju and Kilara, 1998b). The gelation time varies for each process with thermal gelation being the fastest (~30 minutes) and enzymatic gelation the longest (few hours) (Ikeda, 2003; Liang et al., 2020).

This study aims to determine the effect of two emulsification methods followed with two different gelation processes of the dispersed phase in order to produce food-grade microgels from whey proteins. First, experimental parameters for batch and microfluidic emulsification are determined to form droplets with controlled size and polydispersity. Then, two gelation processes, a thermal and an acid gelation are investigated. Finally, microgels prepared with these processes are compared with light microscopy and particle size distribution analysis.

## 2 Materials and Methods

### 2.1 Materials

Whey protein isolate (WPI) powder (Pronativ 95) was purchased from Lactalis (France). It is obtained by membrane filtration and is mainly composed of β-lactoglobulin (≈65%) and α-lactalbumin (≈25%) in their native form.

To prepare protein solutions, the WPI powder was progressively dissolved in milli-Q water (5 wt%) under stirring for at least 2 hours and then stored at 4°C overnight. For solutions submitted to thermal gelation, 200 mM NaCl (Sigma Aldrich, Germany) was added under stirring for 30 min. For acid gelation, solutions were first submitted to a thermal treatment in a water bath (80°C for 30 min) (Moussier et al., 2019b), which induces protein denaturation and partial aggregation. Aggregate size, as estimated by Dynamic Light Scattering (Malvern Nanosizer) is about 130 ± 8 nm.

Sunflower oil was bought from a local supermarket (Cora, France). To stabilize the water-in-oil emulsions, an emulsifier was used: Polyglycerol polyricinoleate (PGPR, ref4150, PALSGAARD) (Nazari et al., 2019) which was dissolved into the oil using magnetic stirring for 2 hours. Different amounts of PGPR were used depending on the conditions, they are indicated wherever necessary.

### 2.2 Microgels preparation

#### 2.2.1 Preparation by batch emulsification

WPI solution (120 mL) was slowly added into 480 mL of sunflower oil solution with 2.5 wt% of PGPR under continuous stirring, in a doubled wall beaker of internal diameter 9 cm (Fig. 1a). An overhead stirrer (Eurostar 60 control, IKA) was used, with a stirring section composed of a three-blade upper stirrer and a lower anchor, as illustrated in the figure. Stirring rate and stirring position play a key role in batch emulsification. Prior experiments were conducted to optimize the stirrer position in the beaker, to avoid the deposit of microgels in the bottom of the vial. Deposit formation was mostly eliminated when the stirrer was in the lowest position, which was selected for this study. In addition, different stirrer configurations were investigated, including one three-blade stirrer, two three-blade stirrers above each other, or one three-blade stirrer above an anchor. The configuration with the anchor was finally selected as it avoided microgel deposit in the bottom of the vial.

The temperature was set using a thermal bath (Heidolph, MR Hei-Standard). For thermal gelation, the temperature was first increased from ambient temperature to 80°C, and then kept at this temperature for 30 min.

For acid gelation, Glucono-delta-lactone (G2164, Sigma Aldrich, Germany), was added at a constant GDL/protein mass ratio (0.08 w/w), immediately before emulsification, and the emulsion was kept at 20°C for 5 hours under stirring. Under these conditions, acidification resulted in a pH value closed to the pHi of β-lactoglobulin, the main component of WPI (Cavallieri and da Cunha, 2008).

#### 2.2.2 Preparation by microfluidics

We developed for this study an home-made microfluidic production system. A hydrophobicaly treated glass microchip with a flow focusing geometry and circular channels (100μm diameter) was bought from Dolomite (US) (Part No.3200434). Both oil (sunflower oil + PGPR) and water (WPI solution) flows were controlled using an *OB1 Mk3+* pressure regulator from Elveflow (France). Both pressures, applied on the continuous oil phase, $P_O$, and on the dispersed aqueous phase, $P_W$ were in the range 0 – 2000 mbar. Droplets formation was observed using a Pixelink D725 camera at a 300 fps framerate.

After production in the microchip, the emulsion flowed into a PTFE tubing immerged in a water-bath. While the residence time in the microchip, $t_1-t_0$ was very short, typically 2 s, the residence time in the thermal bath, $t_2-t_1$, was long enough to ensure droplet gelation, i.e. typically 35 min. At the exit of the thermal bath, droplets were microgels and the suspension was then stored in a small recipient containing oil (1 wt% PGPR) at ambient temperature until the end of the experiment.

The water-bath temperature was fixed at 80°C for thermal gelation and 60°C for the acid gelation.

For acid gelation, GDL (0.08 % w/w) was first dissolved in the protein solution at 4°C. The (protein + GDL) solution was then kept in an ice-cooled water-bath to slow down the acidification process and limit protein aggregation before droplet formation. The residence time of the solution at this temperature, $t_0$, was limited to max 4 hours. Independent measurements at this temperature, over this duration, showed a limited pH decrease, from pH ≈ 6.7 to pH ≈ 5.8, which was not enough to induce protein aggregation, as evidence by viscosity measurement, which did not show any evolution during 4 hours.

#### 2.2.3 From microgel in oil to microgel in water

To observe and characterize microgel suspensions, the particles were transferred in water. The oil (+ PGPR) continuous phase was washed off using centrifugation. The washing consisted of 3 centrifugations at 50G for 20 minutes: After each centrifugation, the supernatant was extracted and replaced by sunflower oil, so that the PGPR was progressively removed from the oil phase. It was followed by 3 centrifugations (same conditions) after which the supernatant was replaced by water, to progressively eliminate the sunflower oil.

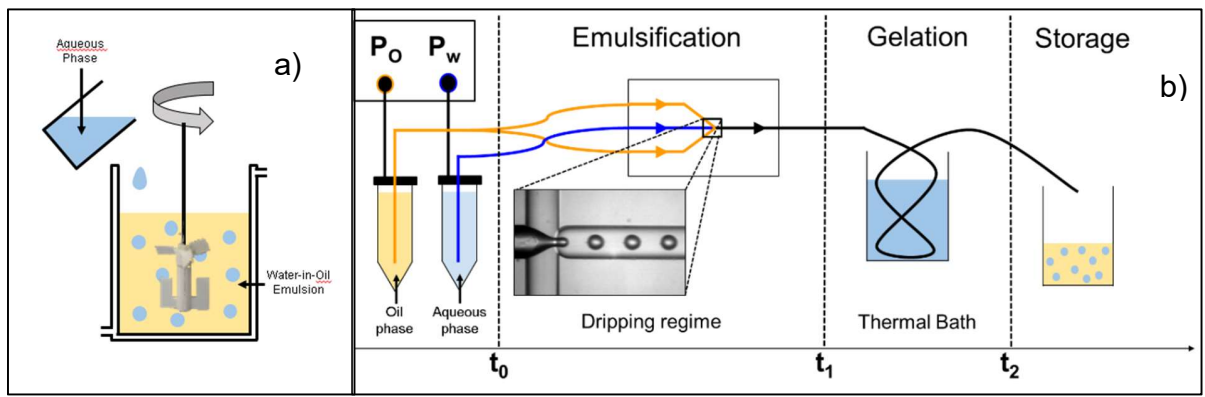

*Figure 1.* Preparation of microgels by (a) batch and (b) microfluidics emulsification. $P_O$ and $P_W$ are the pressure imposed on the Oil and Water solutions, respectively. $t_2$-$t_1$ is the residence time in the tube in which droplets gelation takes place, it is typically in the order of 35 min.

## 2.3 Protein solutions and particles characterization

### 2.3.1 Rheological measurements

Rheological measurements were performed on a stress-controlled rheometer (Anton Paar MCR 301, Graz, Austria) to characterize gel formation at a macroscopic scale.

A concentric cylinder geometry (CC27/S) was used for gelation tracking at a constant strain (γ=0.01) and constant frequency (f=1Hz). For thermal gelation (Th), the solution was first kept at ambient temperature for 15 min before being heated up to 80°C in 8 min, maintained at this temperature for 35 min and then cooled at ambient temperature for 30 min. This procedure reproduced the temperature evolution to which a droplet of aqueous phase was submitted during thermal gelation, either during batch mixing or microfluidic emulsification. For acid gelation, two procedures were followed. In the first one, the temperature was kept at 20°C for 8 hours to mimic the treatment imposed during batch mixing, this procedure being called Batch Acidification (BA). In the second one, a temperature profile that reproduce the temperature evolution to which a droplet was submitted during microfluidic emulsification was imposed. The temperature profile was the following : 15 min at 4°C (temperature in the iced-cooled bath), 5 min at 23°C (temperature in the microchip), 35 min at 60°C (temperature in the water bath), and then 15 min at ambient temperature (temperature of the storage bath); this procedure being called microfluidic acidification (MA).

After complete gelation in the rheometer, the samples were characterized with a frequency sweep, from 100 Hz to 0.01 Hz (at constant strain : γ = 0.01 followed by a strain sweep form 0.0001 to 1 (at constant frequency : f = 1Hz).
All measurements were done in triplicate.

### 2.3.2 pH measurements

The acidification rate was followed with a pH probe (Bioblock scientific). A multi-parameter analyser (Consort C3060) was used to simultaneously follow different experimental conditions. Measurements were done in triplicate.

### 2.3.3 Optical microscopy

Light microscopy was used to observe particles in suspension. A drop of the suspension was placed onto a glass slide with a spacer of 250µm and covered with a cover slip. Samples were observed with a 10X objective lens using a BX 51 microscope (Olympus, Japan).

### 2.3.4 Particle size distribution

The particle size distribution of the microgel samples was acquired from photomicrograph, using the Image Analysis Software ImageJ and a macro developed in our research team for spherical particles diameter analysis. Multiple images from two glass slides for two different samples were analysed, having 500+ particles analysed per sample. A Gaussian curve fit of the size distribution was used to calculate the mean size (µ) and the standard deviation (σ), from which the polydispersity index was determined (pdI = σ/µ).

The particle size distribution of samples obtained from batch emulsification was also acquired using a laser diffraction analyser (Mastersizer 2000, Malvern Instruments). Microgels (refractive index : 1.46) were dispersed in distilled water (refractive index: 1.33). Measurements were made in triplicate

## 3 Results and discussions

### 3.1 Microfluidic emulsification process optimisation

In microfluidic emulsification, 3 regimes of drop generation are observed depending on the relative flow of each liquid phase : dripping, squeezing and jetting (Nunes et al., 2013). In the dripping regime, droplets with a diameter smaller than the microfluidic channel are formed, while in the squeezing regime, droplets are larger and thus squeezed at the wall. In the jetting regime, the dispersed phase flows as a continuous jet in the continuous phase. In a flow-focusing geometry, the capillary number or, for two given phases, the ratio between the dispersed and continuous phase flow, $Q_W/Q_O$, governs the drop generation regime (Cubaud and Mason, 2008; Nunes et al., 2013). In this study, the flow rate of each phase is driven by and proportional to the imposed pressures, so that the pressure ratio $P_W/P_O$ dictates the emulsification regime. The limits of the three regimes were first identified by systematic variations of $P_w$ and $P_O$. For $P_O$ = 1200mbar, Pw is limited to a maximum pressure of 1020mbar. Pressure from 960 to 980 mbar were selected for further investigation of droplet production in the dripping regime.

While droplet formation occurred in the dripping regime, first trials for microgel production resulted in the formation of large and non-spherical microgels (Fig.2, left), suggesting that droplets coalescence occurred after droplets formation and before their gelation.To avoid coalescence, we investigated the effect of an oil-soluble surfactant, PGPR. However, its amount had to be limited, as in flow focusing geometry the oil phase viscosity and the ratio of water to oil phases viscosities affect droplets polydispersity (Nie et al., 2008).

With 0.5 wt% and 0.75 wt% PGPR, coalescence still occured but microgel width and length decreased with the PGPR content, but still remained larger than 100µm, the microfluidic channel diameter. By contrast, with 1wt% PGPR, monodispersed microgels were obtained (Figure 2, right), suggesting coalescence suppression. In the following, this PGPR content is therefore use for microfluidic emulsification.

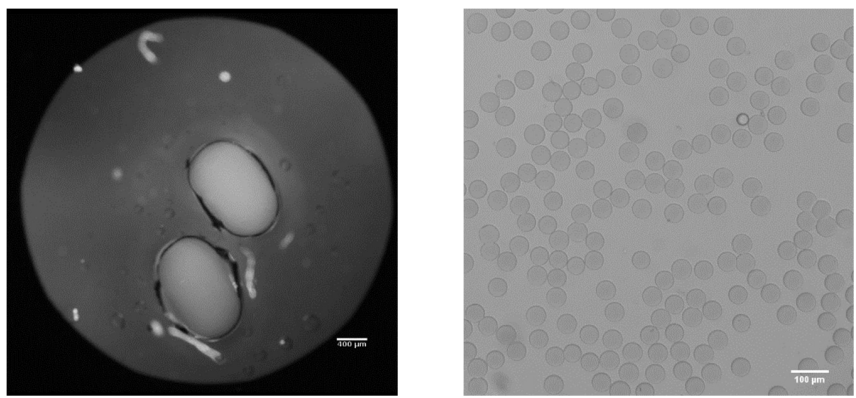

*Figure 2. Gel particles formed without (left) and with 1% PGPR (right) in the oil phase. Scale bars are 400µm (left) and 100 µm (right).*

## 3.2 Protein Gelation

### 3.2.1 Thermal gelation

Rheological experiments were used to characterize on macroscopic gels the gelation process that occurs in droplets. To mimic thermal gelation (Th process), the temperature profile imposed in the rheometer reproduced the temperature evolution to which the protein solution is exposed during microgel preparation, this evolution being the same in batch mixing and in microfluidic emulsification.

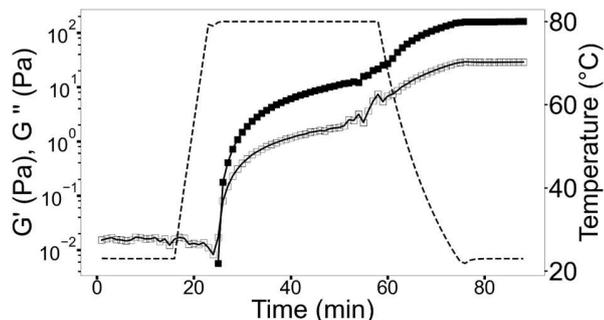

*Figure 3. Evolution of the storage modulus G' (full symbols) and loss modulus G" (empty symbols) during thermal gelation for a solution of 5 wt% WPI. The temperature profile (dashed line) is read on the right axis.*

The protein solution was initially liquid-like, with the storage modulus G' much smaller than the loss modulus G". During heating, G' increased rapidly and exceeded G". The crossover between G' and G" was defined as the gel time: it occurred after 28 min in the conditions presented in Fig. 3. From this point, the system was considered as a gel, still the moduli were increasing till they reached a plateau.

During cooling, G' increased again, reaching a modulus ≈7 times higher than the one reached during heating. This increase is a common feature of protein gels (Basse et al., 2020; Chronakis and Kasapis, 1993) and is considered the result of a reduction of entropy associated with the strengthening of attractive forces (hydrogen bond and Van der Waals forces).

Gelation was achieved once G' reached a plateau after cooling. Here, the plateau was considered reached when the variation of G' was below 1% over 3 minutes, for the case presented in Fig. 3, this occured after 74 min, with G' = 113 Pa.

### 3.2.2 Gelation by acidification

To characterize protein gelation during BA and MA processes, rheological characterization was complemented by pH measurements, in thermal conditions that mimic both processes.

During acidification, the pH continuously decreased at a rate that depends on the temperature, as it can be clearly observed in the MA process: temperature increase up to 60°C was associated with a drastic increase in acidification rates (Fig. 4, right).

As a result, the gelation time (time for wich G'=G") strongly differ between the two processes : it was 196 min and 31 min for BA and MA conditions, respectively. Beside such differences, the pH of gelation (pH at the gel time) were very close: we found pH = 5.5 for BA conditions and pH = 5.6 for MA conditions. The gelation therefore occurred at pH ≈ 5.5, whatever the temperature of the sample. These values are coherent with pH values of 5.8 at gel time which were obtained by others for a same WPI/GDL ratio (Cavallieri and da Cunha, 2008).

For the gel formed in MA conditions, cooling was also associated with an increase in G' by a factor 3.6 for the same reasons as previously described.

The final plateau value was reached after 425 min and 72 min, with G' = 662 Pa and G' = 3130 Pa for BA and MA conditions, respectively. Gels from MA conditions were strongest due to the the cooling step: before cooling, the modulus of MA gels was indeed very close to the one of BA gels, which is coherent to previous results that shown that the temperature of acidification only affects the gelation kinetics, but not the modulus value itself (Kharlamova et al., 2018). Finally, for the same protein concentration, gels obtained by acidification were stiffer than their counterpart obtained by thermal treatment.

### 3.2.3 Gel mechanical properties

Macroscopic gels formed in the rheometer under Th, BA and MA conditions, were characterized under frequency and strain sweep.

Over the full range of frequencies investigated, G' was superior to G" and their low dependency over the frequency confirmed the gel nature of all samples. For the softer samples obtained by thermal gelation, the viscous modulus increases sharply at the highest frequencies. This increases of G" with the frequency is a hallmark of colloidal gel, and has been attributed to the viscous dissipation of the suspending fluid, which is frenquency dependent and, at high frequencies, dominates over the loss modulus of the network (Trappe and Weitz, 2000).

During strain sweep, both G' and G" remained constant until a critical strain at which G' started to decrease. This critical strain value was determined when a variation of G' greater than 10% of it's mean value in the linear domain occured. The strain at the linear domain limit for MA, BA and Th were respectively γ=0.02, γ=0.05 and γ=0.1. For Th gels, after the linear domain, an increase of the shear modulus occured contrarily to MA and BA gels, as previously observed for such structures (Pouzot et al., 2006). Furthermore, for all gelation methods, G' values in the linear domain corresponded with the ones determined during the gelation experiment : 3000 Pa ± 300 Pa, 590 Pa ± 90 Pa, and 130 Pa ± 20 Pa for MA, BA and Th gelation respectively.

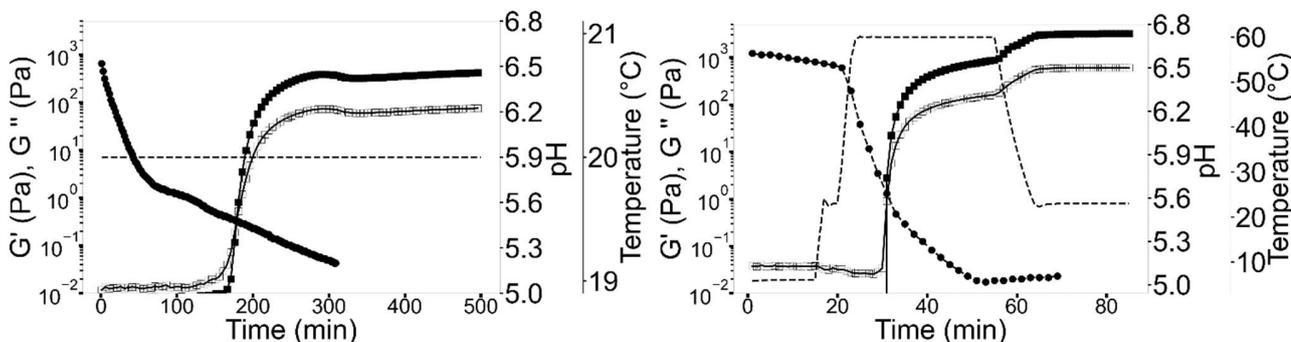

*Figure 4. Evolution of the storage modulus G' (full squares), the loss modulus G" (empty squares) and the pH (full circles) during gelation for a solution of 5 wt% WPI, after GDL addition (0.08 w/w), in BA (left) and MA (right) conditions. The temperature profile (dashed line) is read on the secondary right axis.*

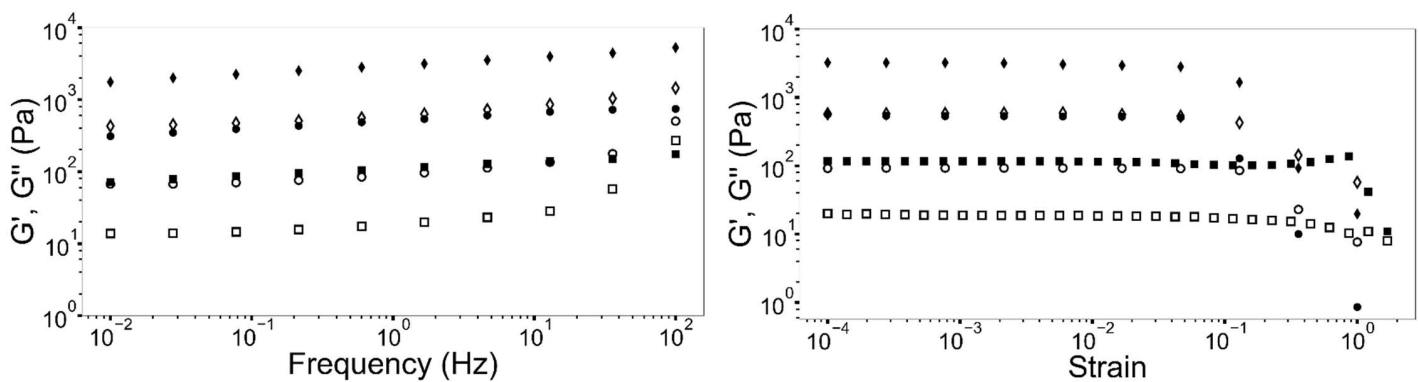

*Figure 5. Evolution of the storage modulus G' (full symbol) and loss modulus G'' (empty symbol if represented) during a frequency sweep (left) and a deformation sweep (right) for a WPI gel from MA (diamond), BA (circle) and TH (square) conditions.*

### 3.3 Microgel observation and size characterization

#### 3.3.1 Microgel suspensions in light microscopy

Depending on the preparation method, different type of particles were obtained, including homogeneous spherical microgels and heterogeneous structures formed by agregation.

Microgels from Th conditions prepared by batch emulsification (Fig 6, top left) irreversibly aggregated during the gelation process. Increasing the PGPR concentration or decreasing heating rate did not avoid this aggregation (data not shown).

Microgels from Th conditions prepared by microfluidic emulsification (Fig 6, top right) also had a tendency to aggregation. In addition, a closer look at micrographs showed that these microgels were not homogeneous, and were characterized by an irregular surface.

Microgels from BA conditions (batch acidification, Fig. 6, bottom left) were globally spherical and most of them were homogeneous, beside in some cases, for the larger ones, and internal core structure could be distinguished from the superficial layer, as illustrated in the inset. Microgels from MA conditions (Fig. 6, bottom right) were spherical and homogeneous.

In conclusion, while microgels obtained via thermal gelation were either inhomogeneous or aggregated, microgels obtained by acidification, either with batch or microfluidic emulsification, were both essentially homogeneous and spherical, and thus considered suitable for further study.

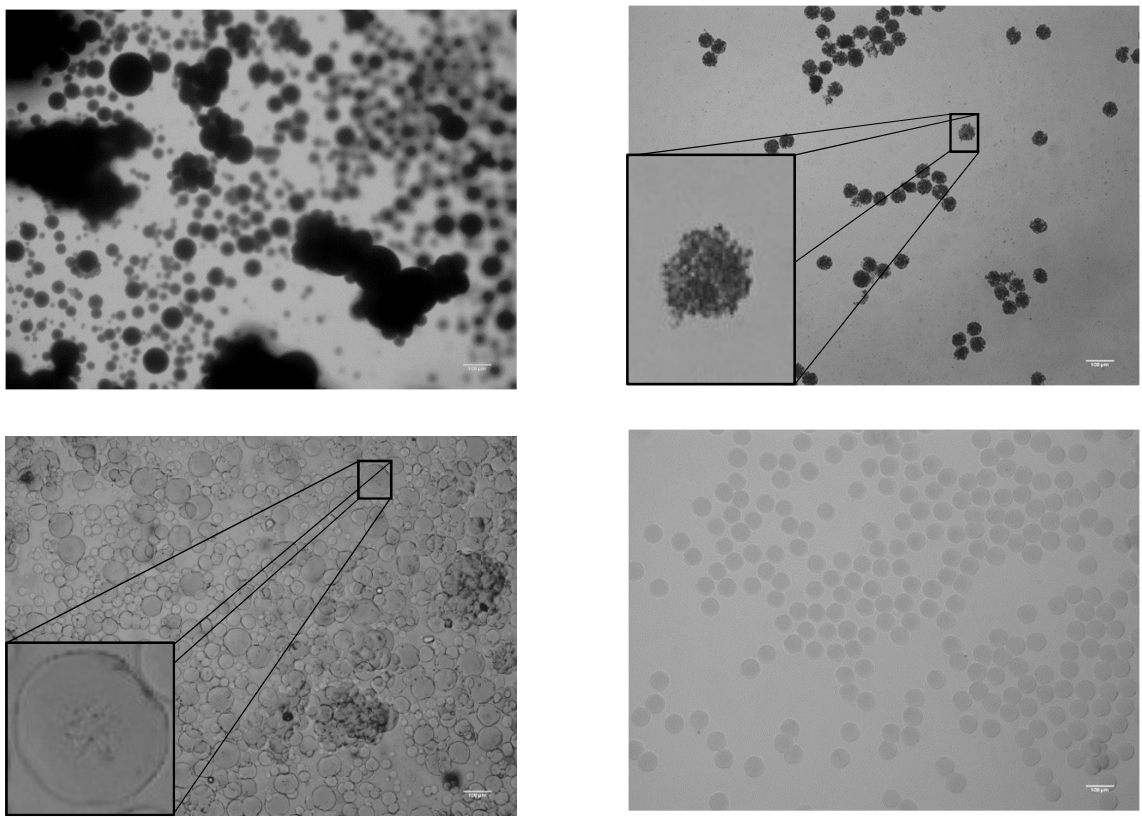

*Figure 6. WPI Microgel suspensions (in water) obtained from microfluidic (right) and batch (left) emulsification using thermal (top) and acid (bottom) gelation. Observation in light microscopy, scale bar is 100 μm.*

### 3.3.2 Influence of the elaboration process on particle size

The size polydispersity of the spherical microgels prepared by acidification was investigated for both BA and MA methods. Furthermore, with the goal of forming particles with adjustable properties, different experimental conditions were studied for microfluidic emulsification, in the dripping regime. To do so, the oil phase pressure $P_O$ was kept constant at 1200 mbar and different water phase pressures $P_W$ (960, 970 and 980 mbar) were imposed.

For the batch emulsification, polydisperse samples were obtained, as shown in Fig. 7. The size distribution determined from image analysis was over a span of 100 µm, with an average diameter of 75µm. For particles prepared by this process, it could be hypothesize that a more monodispersed distribution could be obtained by sieving the suspension or with an higher rotation speed as described previously (Andoyo et al., 2016). We compared this size distribution with the one obtained by laser diffraction (see Insert, Fig. 7); and noticed a significant difference in the measured average diameter : the average diameter determined by laser scattering is much higher (235 µm) than the one determined by image analysis (75 µm). A previous study conducted in our research group already evidenced a difference between the results obtained with the different methods (Moussier et al., 2019a), but of much lower amplitude. Here, we could hypothesize that the difference mainly comes from two reasons. First, the existence of microgels aggregates: in image analysis, only the diameter of individual microgels was determined, while laser scattering only reflects the size of scattering objects, that could be composed of different individual particles aggregated together. Second, in image analysis the measurement of the smaller microgels (<10µm) is limited by the image resolution, therefore they cannot be integrated in the size distribution, and the average diameter determined with this method does not takes them into account.

For microfluidics, results show that tuning the water phase pressure offers the possibility to precisely control the microgels size, while kipping the polydispersity constant.
In addition to a spherical shape, microfluidics also gives microgels that are much more monodisperse than the ones obtained by batch emulsification. This key characteristic, that could already by visually observed in Figure 6, is illustrated in Table 2 by much different values of the polydispersity index resulting from the two emulsification process. Microfluidic emulsification is therefore better suited for the elaboration of spherical, homogeneous and monodispersed microgels.

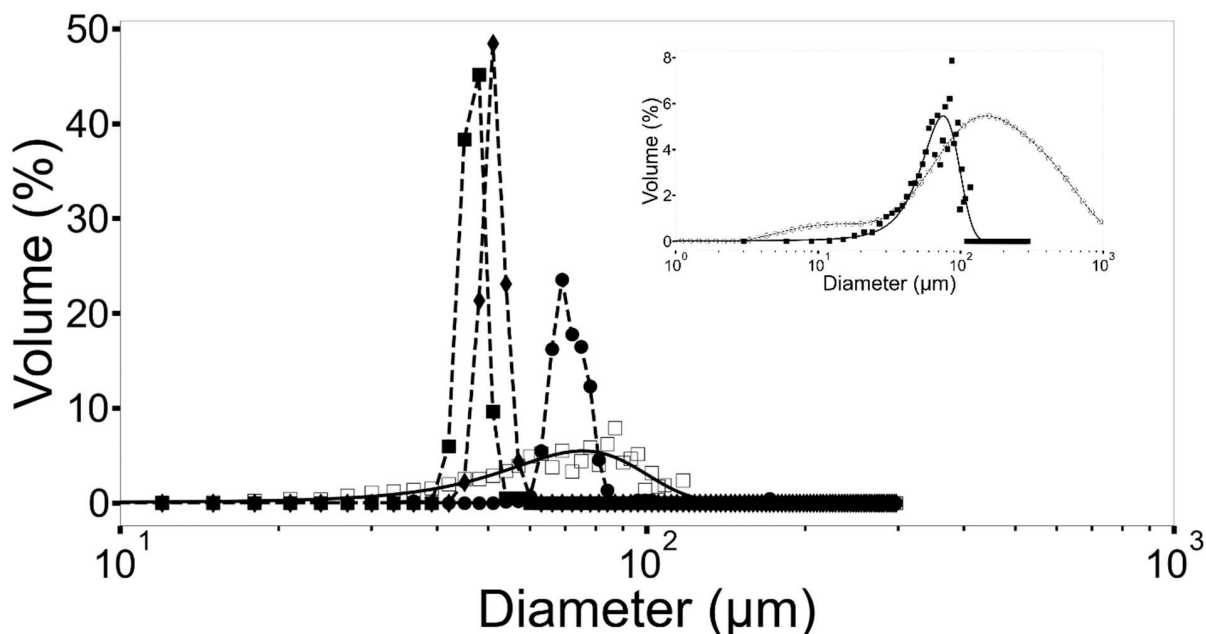

*Figure 7. Particle size distribution for WPI microgels obtained by acidification, the emulsification method being either batch mixing (BA, empty symbols) or microfluidic emulsification (MA, full symbols). Stirring rate used for BA emulsification was 70 rpm (empty squares). Water phase pressures used for MA conditions were 960 mbar (full diamonds), 970 mbar (full circles) and 980 mbar (full squares). Particle size distributions were obtained by image analysis. Insert : comparison between the size distribution obtained by image analysis (full symbols, the continuous line corresponds to the Gaussian fit of the experimental datas) and the one obtained by laser diffraction (empty symbols), for the same BA sample.*

| Emulsification method | Batch Mixing | Microfluidic | | |
|---|---|---|---|---|
| Elaboration conditions | 70 rpm | 960 mbar | 970 mbar | 980 mbar |
| Mean diameter (µm) | 75 | 47 | 51 | 71 |
| Polydispersity index | 0.3 | 0.05 | 0.04 | 0.07 |

*Table 1. Mean size and polydispersity index for acid induced microgels (size distributions were determined from image analysis and presented in Fig. 7).*

## 4 Conclusion

We investigated here the formation of microgels particles from whey protein isolate. Microgels preparation consisted in a two-step process. First, an emulsification step to form spherical droplets (either by batch mixing or microfluidic emulsification) and then a gelation step to form the microgels from the droplets precursors, either by thermal gelation or acidification. We studied the influence of the gelation and emulsification methods on the structure and rheo-physical properties of microgels.

Under the experimental conditions used, thermal gelation resulted in aggregates or heterogeneous microgels. In contrast, gelation by acidification resulted in homogeneous and spherical particles.

Amongst the two emulsification methods, we show that the home-made microfluidics system that we developed on purpose allows the elaboration of more monodispersed spherical microgels independtly of the emulsification conditions ($pdl_{batch} >> 0.1 > pdl_{\mu flu}$). The microgel size can be controlled by the relative flow rate of the oil to the acqueous phase. Furthermore, our results suggest that rheo-physical properties are different for BA ($G' = 590$ Pa) and MA ($G' = 300$ Pa) gels, the higher moduli of the gels obtained in MA conditions being attributed to the higher temperature imposed during acid gelation to speed up the microgel formation in the microfluidic device. These properties could be modified by changing the protein concentration and the temperature imposed during acidification, allowing to tune microgels mechanical properties.

In conclusion, we demonstrate that the set-up that we developed allows the production of a large variery of microgels, which size, surface roughness and mechanical properties can be tuned with the process parameters.

Future work would now be necessary to demonstrate that the mechanical properties of the macroscopic gels are indeed representative of the mechanical properties of their microscopic counterparts, which would required to determined microgel mechanical properties at the particle scale.